\documentclass[12pt,preprint]{iopart}
\usepackage{multirow}
\usepackage{threeparttable}
\usepackage{adjustbox}
\usepackage{graphics}
\usepackage{epstopdf}
\begin{document}
\title{Lifetimes and Oscillator Strengths for Ultraviolet Transitions in Singly-Ionized Tin}
\author{N Heidarian, R E Irving, S R Federman, D G Ellis, S Cheng and L J Curtis}

\address{Department of Physics and Astronomy, University of Toledo, Toledo, OH 43606}
\ead{negar.heidarian@utoledo.edu}

\begin{abstract}
Lifetime measurements using beam foil techniques for the $5s5p^{2}$ $^{2}D_{3/2,5/2}$ levels in Sn~{\sc ii} are presented. The resulting oscillator strengths for transitions at 1699.4, 1831.8 and 1811.2~\AA{} are reported. We also studied these levels with multi-configuration Dirac Hartree-Fock (MCDHF) calculations using a development version of the GRASP2K package. Our experimental and theoretical results are compared with other available studies.

\end{abstract}

\section{INTRODUCTION}

Better understanding of radiative transitions of the Sn$^{+}$ ion has a number of applications including the determination of Sn abundances in the interstellar medium.
The most recent studies on the Sn~{\sc ii} atomic structure for the configurations of our interest were performed by Oliver and Hibbert (2010) using large-scale Breit-Pauli configuration-interaction calculations  and by Alonso-Medina \textit{et al} (2005) with relativistic Hartree-Fock (RHF) calculations. The discrepancies among the available calculations as well as the absence of experimental data especially on the $5s^{2}5d$ $^{2}D$ and $5s5p^{2}$ $^{2}D$ terms was the main motivation of our work. The transitions arising from the $5s^{2}5d$ $^{2}D$ multiplet are of great importance in astrophysics, and due to the strong mixing between the $5s^{2}5d$ $^{2}D_{3/2,5/2}$ and $5s5p^{2}$ $^{2}D_{3/2,5/2}$ levels, more experimental data will lead us to a better understanding of this ion's atomic structure. 
The $5s^{2}5d$ $^{2}D$ has already been experimentally studied by Schectman \textit{et al} (2000) at the University of Toledo using beam-foil techniques and therefore, the focus of the present work is on the $5s5p^{2}$ $^{2}D$ term. The results of our experimental work will be discussed in Section 2 of this paper.
We have also performed a set of calculations with a fully relativistic approach using the GRASP2K package (J{\"{o}}nsson \textit{et al}~2014), with the main focus on the terms mentioned above as well as their radiative transitions to the ground state. These calculations will be discussed in Section 3. Our summary and conclusions are provided in Section 4 of this paper.

\section{EXPERIMENTAL DETAILS}

\subsection{Measurements}

Lifetime measurements were performed using beam foil techniques at the Toledo Heavy Ion Accelerator (THIA; Haar \textit{et al}~1993; Schectman \textit{et al}~2000). Sn$^{+}$ ions were produced by a Danfysik Model 911A Universal Ion Source, initially accelerated at 20 kV and magnetically selected using an analyzing magnet and eventually accelerated to a total kinetic energy of either 200 or 130 keV. The purpose of taking the measurements at two different energies was to study possible systematic effects (see, e.g., Federman \textit{et al}~1992). These Sn$^{+}$ ions were steered toward carbon foils as a solid target using an electrostatic switchyard. The thickness of the foils varied from 2.1 to 2.4 $\mu$g cm$^{-2}$. A beam current of 100 to 110 $nA$ was used to minimize foil breakage. Finally, emission lines were analyzed with an Acton 1m normal incidence vacuum ultraviolet monochromator. Our beam is nearly mono-energetic, and the emission was detected as a function of distance from the target with respect to the slit. By using post-foil velocities, one can achieve a decay curve as a function of time. Depending on the foil thickness, post-foil velocities were estimated to range from 0.5498 to 0.5472 mm ns$^{-1}$ for the 200 keV beam and from 0.4374 to 0.4344 mm ns$^{-1}$ for the 130 keV beam. The precision of the post-foil velocity depends on uncertainties in foil thickness and in the energy calibration of the accelerator. 

Decay curves were taken for lifetime measurements of the $5s5p^{2}$ $^{2}D_{5/2}$ level at 59,463 cm$^{-1}$ using the line at 1811.2~\AA{}. For $5s5p^{2}$ $^{2}D_{3/2}$ at 58,844 cm$^{-1}$, there are two possible branches at 1699.4 and 1831.8~\AA{}, and we used the 1699.4 line since it was relatively brighter. Through analysis of these decay curves, lifetimes were measured. The two levels of interest were quite long lived and beyond the limits usually considered for THIA. Additional measurements to investigate reliability of data were necessary. In both cases, observing the tail of these long lived upper states within one run was not possible due to the limited foil lifetime; therefore, for both lines, we took a set of independent measurements at a fixed distance from the slit with the total dwell time per data point comparable to those for the original decay curve. A variety of fixed distances ranging from close to the slit, to a distance equivalent to about three lifetimes (based on the fitting results of the original decay curve) were chosen and eventually compared with the original decay curve.  After normalization of the extended data points, the decay curve yielded consistent lifetimes as intensity approaching zero at 200 keV. Beam divergence caused by nuclear scattering prevented us from applying this check at 130 keV.

From cascades involving the $4f$ $^{2}D_{7/2}$ level, repopulation of the $5s5p^{2}$ $^{2}D_{5/2}$ level was observed; therefore, two exponentials were necessary to fit this decay curve. Further analysis was required, especially because the possible cascade lifetime was shorter than the primary decay. The method of Arbitrarily Normalized Decay Curves (ANDC; Curtis \textit{et al}~1971) was employed in the analysis, since this method is designed to extract the primary mean lifetime from a decay curve that is affected by cascades. A semi-log plot for the $5s5p^{2}$ $^{2}D_{5/2}$ decay is shown in Figure 1 using the $\lambda1811$ transition. There was no evidence for a noticeable cascade to the $5s5p^{2}$ $^{2}D_{3/2}$ level in our analysis and the lifetime was obtained by means of a single exponential fit, shown in Figure 2 for the $\lambda1699$ transition.

Finally, oscillator strengths for the transitions of interest were derived from our lifetimes. For the $5s5p^{2}$ $^{2}D_{3/2}$ level, there are two possible branches and therefore, branching fractions were needed. We used the branching fractions from our GRASP2K calculations, since they were in better overall agreement with the experimental results compared to the other available values. The branching fractions from the length gauge calculation were used. The percentage difference between the results from the length and velocity gauges for the branching fractions was about 14 percent for both transitions involving this upper level. For the $5s5p^{2}$ $^{2}D_{5/2}$ level, there was only one decay channel and the lifetime yielded the oscillator strength directly.

\subsection{Results}

The results for lifetimes are shown in Table 1 and the resulting oscillator strengths are given in Table 2, along with the previous available experimental and theoretical lifetimes and $f$-values. As mentioned above, the lifetime for the $5s5p^{2}$ $^{2}D_{5/2}$ level using $\lambda1811$ line was obtained with the ANDC analysis; the lifetime of the cascade level was about 3-4~ns with the ANDC analysis and for comparison with other published work, it was 3.2~ns in Alonso-Medina \textit{et al} (2005) calculations and 3.8~ns in the Oliver and Hibbert (2010)  (from quoted transition probabilities and \textit{f}-values).
The weighted average of the lifetimes obtained from the two beam energies is reported, which for the $\lambda1811$ line was obtained from the ANDC analysis and for the $\lambda1699$ line was achieved through a single exponential fit for both energies. For the reported uncertainties, systematic errors (derived from the range in lifetimes from the two energies) were added in quadrature to the weighted uncertainty of the statistical errors associated with the exponential fits.

\section{THEORETICAL CALCULATIONS}

To calculate the energy levels of Sn$^+$ ion with 49 electrons and the ground state of [Kr]$4d^{10}5s^{2}5p$ $^{2}P^{\circ}_{1/2}$ as well as its transition probabilities, a fully relativistic multi-configuration Dirac Hartree-Fock (MCDHF) method was employed using a development version of the GRASP2K package (J{\"{o}}nsson \textit{et al} 2014). From the previous calculations done by others, including the RHF calculations (Alonso-Medina \textit{et al}~2005) and the large-scale Breit-Pauli configuration-interaction calculations (Oliver and Hibbert~2010), the strong mixing between two of the lowest even parity terms, $5s5p^{2}$ $^{2}D$ and $5s^{2}5d$ $^{2}D$ was expected. However, there were inconsistencies among the previous calculations (Alonso-Medina \textit{et al}~2005) and (Oliver and Hibbert~2010), especially for the $5s5p^{2}$ $^{2}D$ multiplet (see Tables 1 and 2). In light of these inconsistencies, as well as the astrophysical importance of these levels, they were the main focus of our calculations. 

We performed separate calculations for odd and even parities for Sn~{\sc ii}. With a core of 46 electrons and 3 valence electrons in the $5s^{2}5p$ shell, we started with one basic reference configuration for odd parities, namely $5s^{2}5p$, which is the lowest odd parity configuration. These orbitals were varied as spectroscopic orbitals to obtain a self-consistent solution. We then started adding more layers of virtual orbitals by generating relativistic configuration state functions (RCSFs) with single and double substitutions from our reference solution, keeping the inner layers fixed and varying the outer layers each time. 

For the even parities, the first eight lowest even parity configurations were included in the reference set. These reference configurations, namely $5s5p^{2}$, $5s^{2}6s$, $5s^{2}5d$, $5s^{2}7s$, $5s^{2}6d$, $5s^{2}8s$, $5s^{2}7d$ and $5s^{2}5g$, were included and varied as spectroscopic orbitals. The purpose of adding all these orbitals was to take care of the possible mixing between $5s5p^{2}$ with $5s^{2}5d$, $5s^{2}6d$ and $5s^{2}7d$; the mixing coefficient of the $5s^{2}8d$ with the $5s5p^{2}$ configuration was negligible compared with the $5s^{2}6d$ and $5s^{2}7d$ and therefore, the $5s^{2}8d$ was not included in the reference set. Again we added outer layers of virtual orbitals one by one using single and double excitation from the reference set to form RCSFs and then varying outer layers keeping the inner layers fixed to obtain the self-consistent solutions. For the even parity solution, a program called RCSFINTERACT included in the GRASP2K package was used to make sure newly formed configuration state functions interact with the reference configurations; this program also reduces the number of generated RCSFs. We then ended the calculations by adding the last layer of orbitals including the $9p$ and $10s$ orbitals for the odd and even parities, respectively (orbitals with $n+l\leq10$) as adding more layers of virtual orbitals at this point did not seem to result in any significant difference in the energy eigenvalues of the levels of interest. These solutions were optimized on the 8 lowest odd terms (16 levels) and 15 lowest even terms (25 levels), but only the reference-configuration orbitals were required to be spectroscopic.

Finally, with 48 relativistic subshells, including 4780 RCSFs for the odd parity solution and 5536 RCSFs for the even parity solution, the Breit interactions were added to the calculations. Using the relativistic configuration interaction (RCI) program also included in the GRASP2K package resulted in the energy eigenvalues shown in Table 3.

The core-valence correlation was also studied by opening the $4d^{10}$ electronic core and allowing one or two electron substitutions from the $4d$ orbital. This step in the calculation did not show any improvement in the energy eigenvalues, and neither were there any significant modifications in the consistency of length and velocity gauges for the oscillator strengths and lifetimes. Therefore, the core-valence correlation calculations were not included in our final results.

As mentioned by several authors before (Oliver and Hibbert 2010) and (Alonso-Medina \textit{et al}~2005), there is a strong interaction between $^{2}D_{3/2,5/2}$ levels of the $5s^{2}5d$ and $5s5p^{2}$ configurations. We noticed that on the basis of the leading expansion coefficient in the LSJ basis, the labels of these levels were switched at different stages of the calculation (due to the strong mixing). In our final energy eigenvalues however, that are reported in Table 3, the labels for these levels were the same as the listed labels for energy values in the National Institute of Standards and Technology (NIST) database. The switching between the labels for these two levels was previously seen and mentioned in Haris \textit{et al} (2014) as well. Despite the ambiguity in the labeling of these levels, the transitions from the lower energy levels to the ground state are the weaker transitions, and the transitions from the higher energy levels are consistently the stronger and more favorable in all the available calculations, including ours. Our experimental measurements support these facts so that the lower energy $^{2}D_{3/2,5/2}$ levels are more long-lived, as opposed to the higher energy $^{2}D_{3/2,5/2}$ levels which have significantly shorter lifetimes. Due to the strong mixing, the dipole matrix elements add up in one case whereas in the other one there are cancellations and this was explained by Oliver and Hibbert (2010) as well. 

Our \textit{ab-initio} theoretical results for the lifetimes and oscillator strengths are shown in Tables 1 and 2, respectively, along with our experimental results and previously published values.

\section{SUMMARY AND CONCLUSIONS}
In this work we presented the results of our lifetime measurements using beam-foil techniques for the $5s5p^{2}$ $^{2}D_{3/2,5/2}$ levels in Sn~{\sc ii} which are presented in Tables 1 and 2. They are the only available experimental measurements for these levels to date. In addition to the values for lifetimes and oscillator strengths included in Tables 1 and 2 from previous work, there was another available calculation by Marcinek and Migdalek (1994). They used a relativistic Hartree-Fock method and included configuration interactions which yielded a multiplet \textit{f}-value of 0.0142 for the $5s^{2}5p$ $^{2}P$ $-$ $5s5p^{2}$ $^{2}D$ transitions and 1.33 for the $5s^{2}5p$ $^{2}P$ $-$ $5s^{2}5d$ $^{2}D$ transitions; from our experimental results we obtained 0.0247 and 1.17, respectively, for the corresponding transitions of these multiplets. We also reported the results of our fully relativistic MCDHF approach for this ion using the GRASP2K package for the \textit{ab-initio} energy eigenvalues of levels of interest in Table 3. Generally, these eigenvalues are in good agreement with the observed energy levels. There is a noticeable discrepancy for the fine structure splitting of the $^{2}P$ terms in our results. Using only the $5s^{2}5p$ $^{2}P$ configuration in the reference set with no substituitions, the value for the fine structure splitting was 4152.74 cm$^{-1}$ which is in good agreement with the observed energy levels; adding virtual orbitals up to the 7p orbital for the odd parity solution only reduced this value to 4079.53 cm$^{-1}$ and lastly by extending the calculations up to the 9p orbital, this value was 3903.18 cm$^{-1}$. However, this difference in the energy eigenvalue of the fine structure did not seem to affect the transition probabilities of levels of interest significantly.

The results of the calculations for the relevant transitions are also reported in length and velocity gauges in Tables 1 and 2. Even though our \textit{ab-initio} calculations showed better consistency between the two gauges for the $5s^{2}5d$ $^{2}D$ term compared to the $5s5p^{2}$ $^{2}D$, they were overall in better agreement with the experimental results compared to other available theoretical calculations, especially for the $5s5p^{2}$ $^{2}D$ term. The velocity gauge results from the GRASP2K calculations are surprisingly in better agreement with the experimental values than the length gauge values. We have shown experimentally that the transitions to the ground state arising from the $5s^{2}5d$ $^{2}D$ term with the higher energies (71,406 and 72,048 cm$^{-1}$) are the more favorable transitions with shorter lifetimes, while the ones from the $5s5p^{2}$ $^{2}D$ term with the lower energies (58,844 and 59,463 cm$^{-1}$) are much longer lived. Previously this was known only theoretically. Our MCDHF calculations support these facts as well.

\ack This work was supported in part by grant HST-AR-12123.001-A from the Space Telescope Science Institute. We thank the anonymous referees for helpful comments that improved the paper.

\section*{References}
\begin{harvard}

\item[] Alonso-Medina A, Col\'{o}n C and  Rivero C  2005 {\it Phys. Scr.} {\bf 71} 154
\item[] Andersen T and Lindgard A 1977 {\it J. Phys. B: At . Mol. Phys.} {\bf 10} 12
\item[] Cardelli J~A, Federman S~R, Lambert D~L and Theodosiou C~E 1993 {\it Astrophys. Lett.} {\bf 416} L41
\item[] Curtis L~J, Berry H~G, and Bromander J 1971 {\it Phys. Lett.} {\bf 34A} 169
\item[] Federman S~R, Beideck D~J, Schectman R~M and York D~G 1992 {\it Astrophys. J.} {\bf 401} 367
\item[] Haar R~R, Beideck D~J, Curtis L~J, Kvale T~J, Sen A, Schectman R M, Stevens H W 1993 {\it Nucl. Instr. Meth. Phys. Res.} {\bf 79} 746
\item[] Haris K, Kramida A and Tauheed A 2014 {\it Phys. Scr.} {\bf 89} 115403
\item[] J{\"{o}}nsson P, Biero\'n J, Brage T, Ekman J, Froese Fischer  C, Gaigalas  G, Godefroid M, Grant I P and Grumer J 2014 {\it A development version of Grasp2K, The Computational Atomic Structure Group, downloaded from: http://ddwap.mah.se/tsjoek/compas/.}
\item[]J{\"{o}}nsson P, Gaigalas  G, Biero\'n J, Froese Fischer  C and Grant I P 2013 {\it Comput. Phys. Commun.} {\bf 184} 2197
\item[] Marcinek R and Migdalek J 1994 {\it  J. Phys. B: At . Mol. Phys.} {\bf 27} 5587 
\item[] Migdalek J 1976 {\it J. Quant. Spectrosc. Radiat. Transfer} {\bf 16} 265
\item[] Oliver P and Hibbert A 2010 {\it J. Phys. B: At . Mol. Phys.} {\bf 43} 074013
\item[] Schectman R~M, Cheng S, Curtis L~J, Federman S R, Fritts M C and Irving R E 2000 {\it Astrophys. J.} {\bf 542} 400

\end{harvard}

\clearpage

\begin{table}[h]
\centering
\begin{threeparttable}
\caption{Lifetimes for $5s^25d$ and $5s5p^{2}$ Levels.}
\label{tab:table}
\begin{indented}
\item[•]
\begin{tabular}{cccccc}
\br 
\footnotesize
 & \multicolumn{4}{c}{$\tau$ (ns)} \\
\cline{2-6} \\
Level & THIA& \multicolumn{2}{c}{GRASP2K\tnote{b}}& Other & Other  \\
\cline{3-4}
 && $\tau_{l}$ & $\tau_{v}$& Experiment  & Theory  \\
\mr 
$5d$ $^2D_{3/2}$ & 0.44$\pm$$0.02$\tnote{a} & 0.38 & 0.40 & 0.50$\pm$$0.05$\tnote{c} &   
 0.41\tnote{d}, 0.45\tnote{e}\\

5$d$ $^2D_{5/2}$ & 0.46$\pm$$0.04$\tnote{a}& 0.43 & 0.44 & $\ldots$ & 0.50\tnote{d}, 0.51\tnote{e}\\

$5s 5p^{2}$ $^{2}D_{3/2}$  & 54.2$\pm$$3.5$\tnote{b} & 33.22 & 52.50 & $\ldots$ &10.31\tnote{d}, 19.30\tnote{e}\\

$5s 5p^{2}$ $^{2}D_{5/2}$ &25.9$\pm$$2.5$\tnote{b}& 21.83 & 30.85 &$\ldots$&11.90\tnote{d}, 15.57\tnote{e}\\

\br
\end{tabular}
\end{indented}
\begin{tablenotes} \footnotesize 
\item[a] Results from the Toledo Heavy Ion Accelerator--Schectman \textit{et al} 2000
\item[b] This work
\item[c] Andersen \& Lindg{\aa}rd 1977 -- Beam-Foil Techniques 
\item[d] Alonso-Medina \textit{et al} 2005 -- Relativistic Hartree Fock with Configuration Interaction (from quoted transition probability)
\item[e] Oliver \& Hibbert 2010 -- Breit-Pauli Configuration Interaction (from quoted oscillator strengths) 

\end{tablenotes}  
\end{threeparttable} 
\end{table}

\begin{table}[h]
\centering
\begin{adjustbox}{width=1\textwidth}
\begin{threeparttable}
\caption{Oscillator Strengths for $5s^25p$ $-$ $5s^25d$ and $5s^25p$ $-$ $5s5p^{2}$ Transitions.}
\label{tab:table}
\begin{indented}
\item[•]

\begin{tabular}{ccccccc}
\br 
\footnotesize
 &  & \multicolumn{4}{c}{$f$-value ($\times10^{-2}$)} \\
\cline{3-7} \\

Transition &  Wavelength &THIA & \multicolumn{2}{c}{GRASP2K\tnote{b}} & Other & Other \\ 
\cline{1-1}\cline{4-5}
{\footnotesize Lower  $-$ Upper} &~\AA{}& &  $f_{l}$ & $f_{v}$&Experiment& Theory \\

\mr
5$p$ $^2P^{\rm o}_{1/2}$ $-$ 5$d$ $^2D_{3/2}$ & 
1400.45 &104$\pm$$5$\tnote{a}& 136.9 & 131.2 & 105$\pm$$10$\tnote{c}&  79.8\tnote{1}, 64\tnote{2}, 125\tnote{3}, 120.5\tnote{4} \\
5$p$ $^2P^{\rm o}_{3/2}$ $-$ 5$d$ $^2D_{5/2}$ & 1475.00 & 106$\pm$$9$\tnote{a} & 111.6 & 107.4 & $\dots$& 74\tnote{1}, 97.0\tnote{3}, 95.24\tnote{4}\\
5$p$ $^2P^{\rm o}_{3/2}$ $-$ 5$d$ $^2D_{3/2}$ & 1489.10 & 17.0$\pm$$1.4$\tnote{a} & 8.0& 7.9 & $\dots$&  8.6\tnote{1}, 10.5\tnote{3}, 5.28\tnote{4}\\ 
\\
5$p$ $^{2}P^{{\rm o}}_{1/2}$ $-$ $5s 5p^{2}$ $^{2}D_{3/2}$ & 1699.42 &0.866$\pm$$0.055$\tnote{b} & 1.47 & 0.81 & $\dots$&5.63\tnote{3}, 2.58\tnote{4}\\
$5p$ $^{2}P^{{\rm o}}_{3/2}$ $-$ $5s 5p^{2}$ $^{2}D_{5/2}$ & $1811.20$& 2.8$\pm$$0.3$\tnote{b}& 3.50 & 2.47 & $\dots$&6.20\tnote{3}, 4.74\tnote{4}\\
$5p$ $^{2}P^{{\rm o}}_{3/2}$ $-$ $5s 5p^{2}$ $^{2}D_{3/2}$ &$1831.76$ &0.425$\pm$$0.027$\tnote{b}& 0.72 & 0.52 & $\dots$&1.61\tnote{3}, 1.11\tnote{4}\\

\br
\end{tabular}
\end{indented}
\begin{tablenotes} \footnotesize 
\item[a] Results from the Toledo Heavy Ion Accelerator--Schectman \textit{et al} 2000
\item[b] This work
\item[c] Andersen \& Lindg{\aa}rd 1977 -- Beam-Foil Techniques
\item[1] Migda{\l}ek 1976 -- Single-Configuration Semiempricial Relativistic Model Potential
\item[2]Cardelli \textit{et al} 1993 -- Coulomb Approximation with Core Polarization
\item[3] Alonso-Medina \textit{et al} 2005 -- Relativistic Hartree Fock (reported in length gauge) 
\item[4] Oliver \& Hibbert 2010 -- Breit-Pauli Configuration Interaction (reported in length gauge)

\end{tablenotes}  
\end{threeparttable} 
\end{adjustbox}
\end{table}

\begin{table}
\centering
\begin{threeparttable}
\caption{Energy Levels of Sn~{\sc ii} in (cm$^{-1}$).}
\label{tab:table}
\begin{indented}
\item[•]
\begin{tabular}{cccc}
\br 
\footnotesize
Configuration &Term & NIST\tnote{a} & GRASP2K \\

\mr 
$5s^{2}5p$ & $^{2}P^{\circ}_{1/2}$ & 0.00 &0.00 \\
$5s^{2}5p$ & $^{2}P^{\circ}_{3/2}$ & 4251.494 & 3903.18 \\ 
\\
$5s5p^{2}$ & $^{4}P_{1/2}$ & 46464.290 & 43038.82 \\
$5s5p^{2}$ & $^{4}P_{3/2}$ & 48368.185 &  44761.84\\
$5s5p^{2}$ & $^{4}P_{5/2}$ & 50730.224 &  47084.17 \\
\\
$5s^{2}6s$ & $^{2}S_{1/2}$ & 56886.363 &  56042.65 \\
\\
$5s5p^{2}$ & $^{2}D_{3/2}$ & 58844.181 & 57652.26 \\
$5s5p^{2}$ & $^{2}D_{5/2}$ & 59463.481 & 58191.17 \\
\\
$5s^{2}5d$ & $^{2}D_{3/2}$ & 71406.142 &  71894.59\\
$5s^{2}5d$ & $^{2}D_{5/2}$ & 72048.260 &  72563.36\\
\\
$5s^{2}6p$ & $^{2}P^{\circ}_{1/2}$ & 71493.273 &  70448.81\\
$5s^{2}6p$ & $^{2}P^{\circ}_{3/2}$ & 72377.4484 &  71258.49  \\
\\
$5s5p^{2}$ & $^{2}S_{1/2}$ & 75954.3 &  78327.50\\
\\
$5s5p^{2}$ & $^{2}P_{1/2}$ & 80455.1 & 85036.56 \\
$5s5p^{2}$ & $^{2}P_{3/2}$ & 81718.3 & 87001.23\\
\\

\br
\end{tabular}
\end{indented}
 \begin{tablenotes} \footnotesize
 \item[a] http://www.nist.gov/pml/data/asd.cfm
 \end{tablenotes}
\end{threeparttable} 
\end{table}

\clearpage

\begin{figure}[ht]
\centering
\includegraphics[width=0.7\textwidth]{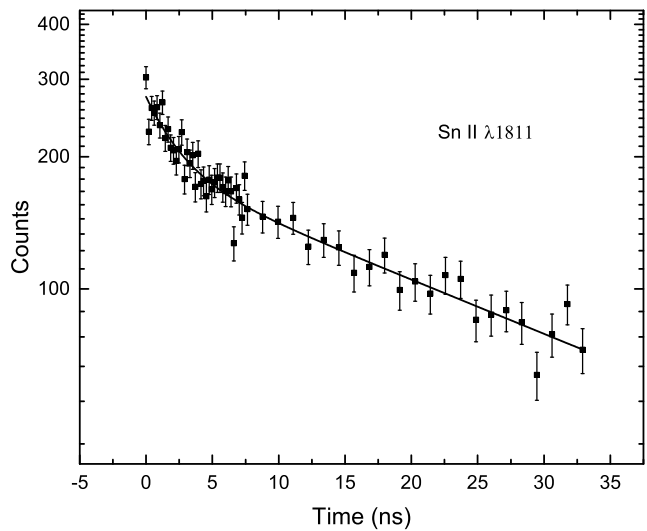}
\caption{Decay curve for the $\lambda1811$ line for a beam energy of 130 keV. The post-foil velocities were used to convert the foil position into a time.}
\end{figure}

\begin{figure}[ht]
\centering
\includegraphics[width=0.7\textwidth]{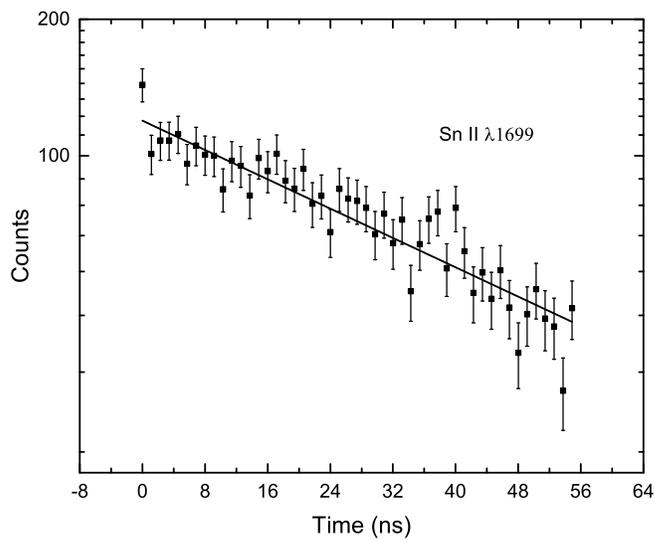}
\caption{Decay curve for the $\lambda1699$ line for a beam energy of 130 keV.}
\end{figure}
\clearpage

\end{document}